# A high-resolution cross-sectional analysis for Fourier-transform scanning tunneling spectroscopy and fully-phased Green-function-based quasiparticle scattering theories

Jhinhwan Lee

Department of Physics, KAIST, 305-701 Daejeon, Korea

The Fourier-transform scanning tunneling spectroscopy (FTSTS) contains rich signals related to the quasiparticle scattering interference and/or the density-wave-like orders which are crucial in interpretation of the ground states and the phase transitions in quasi-2D strongly correlated electron systems such as cuprate and pnictide high-Tc superconductors. The octet model has been used to describe the FTSTS data according to the simplified banana band model of cuprates in superconducting ground state and worked well deep inside the superconducting dome, but away from the superconducting ground state the description starts to show severe limitations, such as appearance of particle-hole asymmetry of the quasiparticle band as will be discussed later. The first efforts to describe the FTSTS signals based on the fully-phased Green's function-based quasiparticle scattering interference theory, however lacked the full similarity to the FTSTS data, in part due to the lack of accurate information on the tunneling and scattering matrix elements. In this paper we discuss how the noise due to the potential disorder and the various artifacts in the raw FTSTS data can be handled by a simple cross-sectional analysis with an isotropic Gaussian averaging and modeling the set-point effect by an energy-independent and position-dependent factor to the density of states function. The result shows that in the cross-sectional presentation an enhanced correlation exists between the FTSTS data and the theoretical simulation even without the accurate model for the tunneling and the scattering matrix elements and without the assumption for particle-hole symmetry, suggesting that it can be a new starting point of a more robust quantum physical analysis framework appropriate for the phase transition study in the high-Tc superconductors and other strongly correlated electron systems.

Recently FTSTS has become an important tool in investigating the momentum space quasiparticle properties of quasi-two-dimensional strongly correlated electron systems such as high-$T_c$ cuprate superconductors. [1,2,3,4,5] The discovery of the dispersing octet peaks in LDOS modulation [1,2] and its theoretical models [6,7] have widened our understanding of the quasiparticles, the superconducting gap and the underlying normal state band structure under various circumstances. When the superconductivity is strong as in nearly optimally doped superconductors at low temperature, all the octet peaks are very well defined with weak background noise-like signals in *q*-space. But as the superconductivity is suppressed by either high temperature, strong magnetic field or non-optimal doping, some of the octet peaks become weaker than the background noise level. To increase the S/N of these octet peaks, various techniques have been developed such as the block reduction averaging, the *q*-space mirror symmetrization based on the symmetry of cuprate lattice, and the p-h quasi-symmetrization (*Z*) based on the p-h anticorrelation of the Bogoliubov quasiparticles with additional important advantage of removing the setpoint factor [3]. But in extreme conditions such as in pseudogap state of the underdoped cuprate superconductor where the existence of Bogoliubov quasiparticles is strongly debated, the efficiency of even the combination of these techniques seems marginal in providing enough S/N and momentum resolution. Here we introduce a different approach that can improve the S/N and the momentum resolution of FTSTS signal drastically and reveal extremely weak non-octet signals contained in FTSTS whose strength becomes comparable to octet signals even in conditions of very weak superconductivity.

The energy-independent noise-like patterns in *q*-space is originating from the random distribution of scattering centers and its suppression by block averaging is demonstrated in [6]. Here we propose to use isotropic gaussian averaging (IGA) which has the following additional advantages over block averaging: (1) it has an exponentially fast momentum space cut-off instead of the slow Sinc function wise roll off of block averaging. (2) We don't lose any pixel resolution unlike block reduction averaging. (3) Center of weight of an arbitrary shaped peak does not shift after the averaging. (4) Compared to the traditional transverse (1D) averaging (averaged over ~2*r* pixels where *r* is the half width) often used in FTSTS line cuts, isotropic (2D) gaussian averaging averages more pixels (averaged over ~$\pi r^2$ pixels) in the same effective radius *r*, as is shown in FIG 1. The choice of the radius *r* can be made by monitoring the S/N in FTSTS line cuts but usually the finest feature radius in the original FTSTS can be chosen as *r* to achieve enough S/N and minimal signal broadening as demonstrated in FIG 2. In our typical 220X220 pixel FTSTS images covering a *q*-space square formed by *q*=(±2.2π,0) and *q*=(0,±2.2π) corners, we chose radius of 4 pixels which will be used throughout this paper. Mathematically, isotropic gaussian averaging with radius *r* (denoted IGA*r*) is achieved by Fourier-transform-based convolution:

$$(f \otimes K_r)(q_x, g_y) = FT^{-1}\{FT\{f(q_x, g_y)\}FT\{K_r(q_x, q_y)\}\} \tag{1}$$

where $f(q_x,q_y)$ is the *q*-space FTSTS image at each energy and $K_r(q_x,q_y)$ is a 220x220 pixel image of normalized gaussian with radius *r* (whose pixel values sum up to 1) used as a convolution kernel.

Since the absolute value of the noise signal is averaged in IGA, its background level contains offset proportional to the noise level. To enhance the contrast of IGA further by removing the background, we can optionally subtract a heavily IGA broadened image (IGA15) from the IGA4 image with a multiplication factor *c* between 0 and 1. The multiplication factor *c* is chosen such that the resulting image shows maximum contrast.  For experimental FTSTS data we chose $c \sim 0.75$ and the net process (denoted IGA4-0.75X15) is equivalent to applying a single convolution kernel:

$$K(q_x, q_y) = Ae^{-(q_x^2+q_y^2)/a^2} - cBe^{-(q_x^2+q_y^2)/b^2} \qquad (2)$$

where $A$ and $B$ are the normalization factor for each gaussian and $a$=4 pixels, $b$=15 pixels and $c$=0.75 for all the experimental FTSTS data shown in this paper.  For noiseless simulated FTSTS data to be discussed later, we can simply choose $c$=0.  The convolution kernel IGA4-0.75X15 in $q$-space and its Fourier space is depicted in FIG 3 (a) and (b).  In case we only need to trace the top negative curvature part (head) of each peak, we can alternatively take a minus Laplacian ($-\nabla^2$) of IGA$a'$ (denoted NL-IGA$a'$) and set the negative value region as zero (or black color in this paper). The convolution kernel NL-IGA7 in $q$-space and its Fourier space is depicted in FIG 3 (c) and (d).  Note that weak peaks near a strong peak become invisible in case of NL-IGA$a'$ (FIG 3 (i)) and IGA$a$-$c$X$b$ with $c \sim 1$ (FIG 3 (h)).  These two kinds of background removal process are only for contrast enhancement and should be used with care to analyze qualitative trends (the existence and dispersion) of fine features in FTSTS line cuts.  We should go back to the IGA or unfiltered data whenever more quantitative analysis of amplitude is necessary.  The overall effects of IGA$a$-$c$X$b$ and NL-IGA$a'$ are best demonstrated in FTSTS line cuts of UD20 Bi2212 taken at T=4K shown in FIG 3 (j)-(o).  Note the multiple peaks in $q_7$ region are resolved in (n) and (o).  Also note that the $q_5$ consists of at least two heterogeneous peaks with different dispersion characteristics as shown in (k) and (l).

Application of this technique to underdoped Bi2212 reveals plenty of fine structures and non-octet peaks whose origins need to be clarified for accurate QPI analysis in weak superconductivity condition.  These signals can be classified into three cartegories: (1) Set-point (1st order $q$-vector mixing) artifact in $g(q,E)$ and $D(q,E)$=FT{$g(r,E)$-$g(r,-E)$}, (2) 2nd order $q$-vector mixing artifact in $Z(q,E)$=FT{$g(r,E)/g(r,-E)$}, (3) Non-octet fine-structures predicted from FTSTS simulation.

### 1. Set-point (1st order $q$-vector mixing) artifacts in $g(q,E)$ and $D(q,E)$

In general, the differential conductance ($g$) measured in FTSTS and its anti-symmetrized quantity ($D$) can be written as:

$$g(\mathbf{r}, E) = \frac{eI_0 N(\mathbf{r}, E)}{\int_0^{eV_0} N(\mathbf{r}, E) dE} = f(\mathbf{r}) \cdot N(\mathbf{r}, E) \qquad (3)$$

$$f(\mathbf{r}) \equiv \frac{eI_0}{\int_0^{eV_0} N(\mathbf{r}, E) dE} \qquad (4)$$

$$D(\mathbf{r}, E) \equiv g(\mathbf{r}, E) - g(\mathbf{r}, -E) = f(\mathbf{r}) \cdot (N(\mathbf{r}, E) - N(\mathbf{r}, -E)) \qquad (5)$$

where $V_0$ is the set-point voltage, $I_0$ is the set-point current, $N(r,E)$ is the sample LDOS accessible at the tip apex (tunneling matrix element effect included) and $f(r)$ is the set-point factor with spatial dependence only. $N(r,E)$ and $f(r)$ can be Fourier expanded as:

$$N(\mathbf{r},E) = a_0(E) + \sum_i a_i(E)\cos(\mathbf{q}_i(E)\cdot\mathbf{r} + \phi_i(E)) \tag{6}$$

$$f(\mathbf{r}) = c_0 + \sum_t c_t \cos(\mathbf{q}_t \cdot \mathbf{r} + \phi_t) \tag{7}$$

$$\begin{aligned}
D(\mathbf{r},E) &\equiv g(\mathbf{r},E) - g(\mathbf{r},-E) = f(\mathbf{r})\cdot(N(\mathbf{r},E) - N(\mathbf{r},-E)) \\
&= \left(c_0 + \sum_t c_t \cos(\mathbf{q}_t\cdot\mathbf{r} + \phi_t)\right) \times \\
&\quad \begin{pmatrix} a_0(E) + \sum_i a_i(E)\cos(\mathbf{q}_i(E)\cdot\mathbf{r} + \phi_i(E)) \\ -a_0(-E) - \sum_i a_i(-E)\cos(\mathbf{q}_i(-E)\cdot\mathbf{r} + \phi_i(-E)) \end{pmatrix}
\end{aligned} \tag{8}$$

In general $\mathbf{q}_i$ and $\mathbf{q}_t$ correspond to every pixel in FTSTS image. But we can assume, to the lowest order, that $\mathbf{q}_t$ contains only the peaks for modulations that survive the LDOS integration over the wide setpoint bias voltage range, which are mostly the Bragg peaks and the supermodulation peaks, and that $\mathbf{q}_i$ covers all the positions for the 32 octet peaks as well as the atomic peaks and the supermodulation peaks that show up in experimental Z map defined in the section 2 below.

Therefore from the trigonometric rule, $\cos A \cos B = \frac{1}{2}(\cos(A+B) + \cos(A-B))$, it is straightforward to show that $g(\mathbf{r},E)$ and $D(\mathbf{r},E)$ contain the following wave vectors with corresponding amplitudes:

$g(\mathbf{r},E)$:
$$\begin{aligned}
&(i)\{\mathbf{q} = \mathbf{q}_i(E), A = c_0 \cdot a_i(E)\}\text{ (physical signals)}, \\
&(ii)\{\mathbf{q} = \mathbf{q}_t, A = c_t \cdot a_0(E)\}\text{ (setpoint - artifact signals)}, \\
&(iii)\{\mathbf{q} = \mathbf{q}_t + \mathbf{q}_i(E), A = c_t \cdot a_i(E)\}\text{ (interference of the signals }(i)\text{ and }(ii))
\end{aligned} \tag{9}$$

$D(\mathbf{r},E)$:
$$\begin{aligned}
&(iv)\{\mathbf{q} = \mathbf{q}_i(E), A = c_0 \cdot a_i(E)\}\text{ (physical signals)}, \\
&(v)\{\mathbf{q} = \mathbf{q}_i(-E), A = -c_0 \cdot a_i(-E)\}\text{ (physical signals)}, \\
&(vi)\{\mathbf{q} = \mathbf{q}_t, A = c_t \cdot (a_0(E) - a_0(-E))\}\text{ (setpoint - artifact signals)}, \\
&(vii)\{\mathbf{q} = \mathbf{q}_t \pm \mathbf{q}_i(E), A = \tfrac{1}{2}c_t a_i(E)\}\text{ (interference of the signals }(iv)\text{ and }(vi)), \\
&(viii)\{\mathbf{q} = \mathbf{q}_t \pm \mathbf{q}_i(-E), A = -\tfrac{1}{2}c_t a_i(-E)\}\text{ (interference of the signals }(v)\text{ and }(vi))
\end{aligned} \tag{10}$$

In FIG 4 (a)-(c), the interference alias peaks of the 32 QPI peaks due to the Bragg peaks at $\mathbf{q}_t=(\pm 2\pi, 0)$ and $\mathbf{q}_t=(0,\pm 2\pi)$ are marked by squares and those due to the supermodulation peaks are marked by crosses to show how strong the corresponding interference signals are in $g$, $D$ and $Z$ maps.

## 2. 2nd order $q$-vector mixing artifacts in $Z(q,E)$

Recently a dimensionless quantity $Z$ was introduced to eliminate the set-point factor and also enhance the p-h antisymmetric Bogoliubov QPI signal for the cuprate superconductors [3].

$$Z(\mathbf{r},E) \equiv \frac{g(\mathbf{r},E)}{g(\mathbf{r},-E)} = \frac{N(\mathbf{r},E)}{N(\mathbf{r},-E)} \qquad (11)$$

However, *Z(r,E)* can also contain non-negligible artifact similar to *D(r,E)* as described below.

$$\begin{aligned}
Z(\mathbf{r},E) &= \frac{N(\mathbf{r},E)}{N(\mathbf{r},-E)} \\
&= \frac{a_0(E)}{a_0(-E)} \frac{1+\sum_i b_i(\mathbf{r},E)}{1+\sum_i b_i(\mathbf{r},-E)} \quad \left[b_i(\mathbf{r},E) \equiv \frac{a_i(E)}{a_0(E)} \cos(\mathbf{q}_i(E)\cdot\mathbf{r}+\phi_i(E))\right] \\
&= \frac{a_0(E)}{a_0(-E)} \left(1+\sum_i b_i(r,E)\right) \times \left(1-\left(\sum_i b_i(r,-E)\right)+\left(\sum_i b_i(r,-E)\right)^2 -\cdots\right) \\
&= \frac{a_0(E)}{a_0(-E)} \left(1+\sum_i b_i(r,E)-\sum_i b_i(r,-E)-\sum_i\sum_j b_i(r,E)b_j(r,-E)+\left(\sum_i b_i(r,-E)\right)^2 -\cdots\right) \\
&= \frac{a_0(E)}{a_0(-E)} + \sum_i \frac{a_i(E)}{a_0(-E)} \cos(\mathbf{q}_i(E)\cdot\mathbf{r}+\phi_i(E)) - \sum_i \frac{a_0(E)a_i(-E)}{a_0(-E)^2} \cos(\mathbf{q}_i(-E)\cdot\mathbf{r}+\phi_i(-E)) \\
&\quad -\sum_i\sum_j \frac{a_i(E)a_j(-E)}{a_0(-E)^2} \cos(\mathbf{q}_i(E)\cdot\mathbf{r}+\phi_i(E))\cos(\mathbf{q}_j(-E)\cdot\mathbf{r}+\phi_j(-E)) \\
&\quad +\sum_i\sum_j \frac{a_0(E)a_i(-E)a_j(-E)}{a_0(-E)^3} \cos(\mathbf{q}_i(-E)\cdot\mathbf{r}+\phi_i(-E))\cos(\mathbf{q}_j(-E)\cdot\mathbf{r}+\phi_j(-E)) + \cdots
\end{aligned} \qquad (12)$$

Therefore *Z(**r**,E)* contains the following wave vectors with corresponding amplitude:

$$\begin{aligned}
&Z(\mathbf{r},E): \\
&(i)\{\mathbf{q}=\mathbf{q}_i(E), A=a_i(E)/a_0(-E)\} \text{ (physical signals)}, \\
&(ii)\{\mathbf{q}=\mathbf{q}_i(-E), A=-a_0(E)a_i(-E)/a_0(-E)^2\} \text{ (physical signals)}, \\
&(iii)\{\mathbf{q}=\mathbf{q}_i(E)\pm\mathbf{q}_j(-E), A=-a_i(E)a_j(-E)/2a_0(-E)^2\} \text{ (interference of the signals } (i) \text{ and } (ii)), \\
&(iv)\{\mathbf{q}=\mathbf{q}_i(-E)\pm\mathbf{q}_j(-E), A=a_0(E)a_i(-E)a_j(-E)/2a_0(-E)^3\} \text{ (interference of the two signals in } (ii)) \\
&(v)\{\mathbf{q}=\text{all possible multiple combinations of } \mathbf{q}_i(E), \mathbf{q}_i(-E), \mathbf{q}_j(-E), \cdots, \mathbf{q}_k(-E)\} \\
&\quad \text{(interference of the signals in } (ii) \text{ and } (iii))
\end{aligned} \qquad (13)$$

Unlike *D(r,E)*, the modulations (*i*) and (*ii*) have energy-dependent amplitude weight of order $\sim 1/a_0(E)$ which is enhanced strongly at low energy for a typical gapped average spectrum. *Z(r,E)* also contains infinitely many weak modulations with all possible combinations of *$q_i$* vectors and they become noticeable when one of the interfering peaks has a very well defined *q* vector such as the Bragg peaks or the supermodulation peaks. This 2[nd] order *q*-vector mixing in *Z* can explain a part of the heterogeneous peak with intensity proportional to the Bragg peak intensity appearing at $q=2\pi-q_1$ in FIG 4 (m)&(o) and FIG S1,S2. As can be seen in FIG 10, a heterogeneous peak is also expected in a pure d-wave superconductivity Green function simulation, however, due to this complication, $q_5$ is not a very good peak to assign physical meaning and was excluded from the octet analysis in [4].

Another useful feature of cross-sectional FTSTS analysis with IGA is to reconfirm the octet model as demonstrated in FIG.5. In the octet model, $q_4$ is twice the *k* vector of the highest density of states position (the ends of the 'bananas') in *k*-space, and the $q_1$ and $q_5$ vectors are the projections of $q_4$ onto (0,0,)-(0,2π) axis, and the $q_7$ vector is the projection of $q_4$ onto (0,0,)-(π,π) axis as shown in FIG.5 (a) and (b). Therefore if we take the high-symmetry-axis projections of $q_4$ as we take the curvilinear line cut along the curve on which $q_4$ is dispersing, we should be able to reproduce the dispersions of the $q_1$, $q_5$ and $q_7$ peaks. This is demonstrated in FIG.5 (c)-(g) where the dispersions of $q_1$, $q_5$ and $q_7$ are reproduced well from the projection of the $q_4$ peak dispersion, confirming the overall validity of the octet model.

### 3. Non-octet fine-structures predicted from quasiparticle interference simulation

It has been predicted in the early QPI simulations for cuprates [6,7] that plenty of non-octet fine structures originate even from the simplest *d*-wave superconductivity Green function scattering simulations and in the weak scattering approximation off a single scattering impurity, as reproduced in (b) and (d) of FIG 6,7.

In order to compare how well those fine features in theoretical simulations agree with those in experiments, we followed the simulation scheme described in [6] as described below.

The local density of states is given by the spin-trace of the imaginary part of the 'bare' Green function

$$\mathbf{G}_0(\mathbf{k},\omega) = ((\omega+i\delta)\mathbf{I} - \varepsilon_\mathbf{k}\boldsymbol{\sigma}_3 - \Delta_\mathbf{k}\boldsymbol{\sigma}_1)^{-1} = \begin{pmatrix} \omega+i\delta-\varepsilon_\mathbf{k} & -\Delta_\mathbf{k} \\ -\Delta_\mathbf{k} & \omega+i\delta+\varepsilon_\mathbf{k} \end{pmatrix}^{-1} \quad (14)$$

integrated over the Brillouin zone:

$$N_0(\omega = E) = \frac{1}{\pi}\int \frac{d^2\mathbf{k}}{(2\pi)^2}\operatorname{Im}\{Tr\mathbf{G}_0(\mathbf{k},\omega)\}. \quad (15)$$

The local density of states without scattering process shows no position dependence as expected from the case without the presence of any density-wave-like order [8].

The QPI modulation in the local density of states is then given by the same operation on the 'dressed' portion of the full Green function

$$\mathbf{G}(\mathbf{k},\omega) = \mathbf{G}_0(\mathbf{k},\omega) + \mathbf{G}_0(\mathbf{k},\omega)\cdot\mathbf{T}(\mathbf{k}+\mathbf{q},\mathbf{k};\omega)\cdot\mathbf{G}_0(\mathbf{k},\omega) \quad (16)$$

where the dressing occurs due to the multiple scattering off the various scattering potential disorders as the 'bare' quasiparticle propagates, whose effect is included in the following simplified scattering matrix.

$$\mathbf{T}(\mathbf{k}+\mathbf{q},\mathbf{k};\omega) \sim \mathbf{T}(\omega) = \left( (V_s\boldsymbol{\sigma}_3 + V_m\mathbf{I})^{-1} - \int \frac{d^2\mathbf{k}}{(2\pi)^2} \mathbf{G}_0(\mathbf{k},\omega) \right)^{-1}$$

$$= \left( \begin{pmatrix} V_m + V_s & 0 \\ 0 & V_m - V_s \end{pmatrix}^{-1} - \int \frac{d^2\mathbf{k}}{(2\pi)^2} \begin{pmatrix} \omega + i\delta - \varepsilon_\mathbf{k} & -\Delta_\mathbf{k} \\ -\Delta_\mathbf{k} & \omega + i\delta + \varepsilon_\mathbf{k} \end{pmatrix}^{-1} \right)^{-1} \quad (17)$$

Here we assumed that the scattering strength is independent of momentum which may be the case for scattering off a single impurity at the origin or completely disordered impurities without any preferred length scale.

The total density of states is then given by

$$N(\mathbf{q},\omega) = \frac{1}{\pi} \int \frac{d^2\mathbf{k}}{(2\pi)^2} \mathrm{Im}\{Tr\mathbf{G}(\mathbf{k},\omega)\}$$
$$= N_0(\omega) + \tilde{N}(\mathbf{q},\omega) \quad (18)$$

where the density of states modulation due to scattering is defined as

$$\tilde{N}(\mathbf{q},\omega) = -\frac{1}{2\pi i}\left[ A_{11}(\mathbf{q},\omega) + A_{22}(\mathbf{q},-\omega) - A_{11}^*(-\mathbf{q},\omega) - A_{22}^*(-\mathbf{q},-\omega) \right] \quad (19)$$

Here the modulation amplitude function is defined and calculated as

$$\mathbf{A}(\mathbf{q},\omega) = \int_{\substack{\mathbf{k}\in BZ \\ \mathbf{k}+\mathbf{q}\in BZ}} \frac{d^2\mathbf{k}}{(2\pi)^2} \mathbf{G}_0(\mathbf{k}+\mathbf{q},\omega) \cdot \mathbf{T}(\mathbf{k}+\mathbf{q},\mathbf{k};\omega) \cdot \mathbf{G}_0(\mathbf{k},\omega). \quad (20)$$

Once the modulation part of the density of states is calculated, the simulated FTSTS data corresponding to *Z* map or *D* map can be produced just by the *D* map process

$$Z(\mathbf{q},\omega) \sim D(\mathbf{q},\omega) = g(\mathbf{q},\omega) - g(\mathbf{q},-\omega) \sim N(\mathbf{q},\omega) - N(\mathbf{q},-\omega) \sim \tilde{N}(\mathbf{q},\omega) - \tilde{N}(\mathbf{q},-\omega) \quad (21)$$

since the theoretical density of states is free from the setpoint artifact.

As we go through each FTSTS simulation, we adjusted the five parameters $\Delta_0$, $t_0$, $\delta$, $V_s$ and $V_m$ so that we have maximum similarities with each experimental FTSTS data in all the QPI peaks' dispersions along the high-symmetry axes in cross-sectional mode. We first determined $\Delta_0$ from the slope of the $q_7$ dispersion and then determined $t_0$ from the locations of other ($q_1$, $q_3$ and $q_5$) peaks. The infinitesimal broadening factor $\delta$ is chosen as a small fixed value of 6 meV. For the UD74 Bi2212 samples, we obtained $\Delta_0$=55 meV, $t_0$=71.5 meV, and tried two different kinds of scattering impurities: (1) non-magnetic scattering ($V_s$=100 meV, $V_m$=0 meV) and (2) magnetic scattering ($V_s$=0 meV, $V_m$=100 meV). The comparison between the experiment and the simulation is shown in FIG 6,7 in ordinary 2D *q*-space representation and in FIG.8 in cross-sectional representation. In FIG.8 the top row is the experimental FTSTS line cuts using the IGA applied and the second and third rows are the cases of non-magnetic and magnetic scattering only respectively with the same amount of IGA applied. Note that there are fair similarities between the experiment and the simplest *d*-wave superconductivity Green function simulation in terms of the existence and dispersion of the fine features for both magnetic and non-magnetic scattering matrices, even though the intensity distributions are noticeably different possibly due to the unknown exact functional form of the scattering and tunneling matrix element. Apparently this lack of similarity in intensity distribution is what made the direct comparison of the experimental data and the theoretical simulation data so difficult in the ordinary 2D *q*-space representation as demonstrated in FIG 6,7.

We can apply this method to cuprates with different doping and temperature, and possibly other physical parameter such as applied magnetic field and various number of coupled copper-oxide layers within a unit cell and so on. The method is demonstrated in FIG 9 and 10 on a data published in [4] taken on UD37 Bi2212 sample as we vary the temperature from 4K to 55K. Apparently we can explain the 4K data fairly well with the scheme following the reference [6] which assumed perfect *d*-wave superconductivity only. If we look at the trend of the dispersion of $q_1$ of $g(q,E)$ in FIG 9 as $T$ varies, it is obvious that the particle-hole symmetry is weakened and the dispersion slowly approaches that of a normal state dispersion of cuprate (i.e. dispersions in the positive and negative energies approaching a single straight line) as $T$ increases through $T_c$.

In summary we have shown that, due to the enhanced S/N of the cross-sectional FTSTS analysis with IGA that can resolve fine physical features and possible artifacts, the FTSTS line cut comparison in high symmetry directions can be a powerful analysis method in testing new fully phased Green's function based theories of an unknown state of a new quasi-2D strongly correlated electron material with plenty of information to fit theoretical parameters to, even in situations with very limited available information on the scattering and tunneling matrix elements. Of course we need to maintain sufficient spatial (256 x 256 pixels over 46 nm x 46 nm) and energy (35 layers with 2mV energy step) resolution, wide temperature range (from 4K up to $2T_{c,max}$) and uniform and ideal tip quality for such a comparison to be meaningful. When a phase-diagram-wide FTSTS measurement technique with the above criteria is developed and when a fully-phased nonzero-temperature Green's function analysis with a proper *T*-dependent gap function, properly determined by a self-consistent quantum field theory that can handle the BCS-like divergence, is developed, the analysis technique shown here will have a potential to uniquely determine the exact physics of all part of the phase diagram of the cuprate and other quasi-2D strongly correlated electron systems accessible by spectroscopic-imaging STM at the most fundamental many-body quantum physical level.

**Acknowledgement**

The author is thankful to J.C. Davis for access to facilities and use of data.  The author is also thankful to J.C. Davis, K. Fujita, A. Schmidt, C.K. Kim, Jinho Lee, D.-H. Lee, S. Uchida, K. Levin, C.Y. Kim, T. Park, K. Park, T. Hanaguri, E.-A. Kim and M. Lawler for helpful discussions.

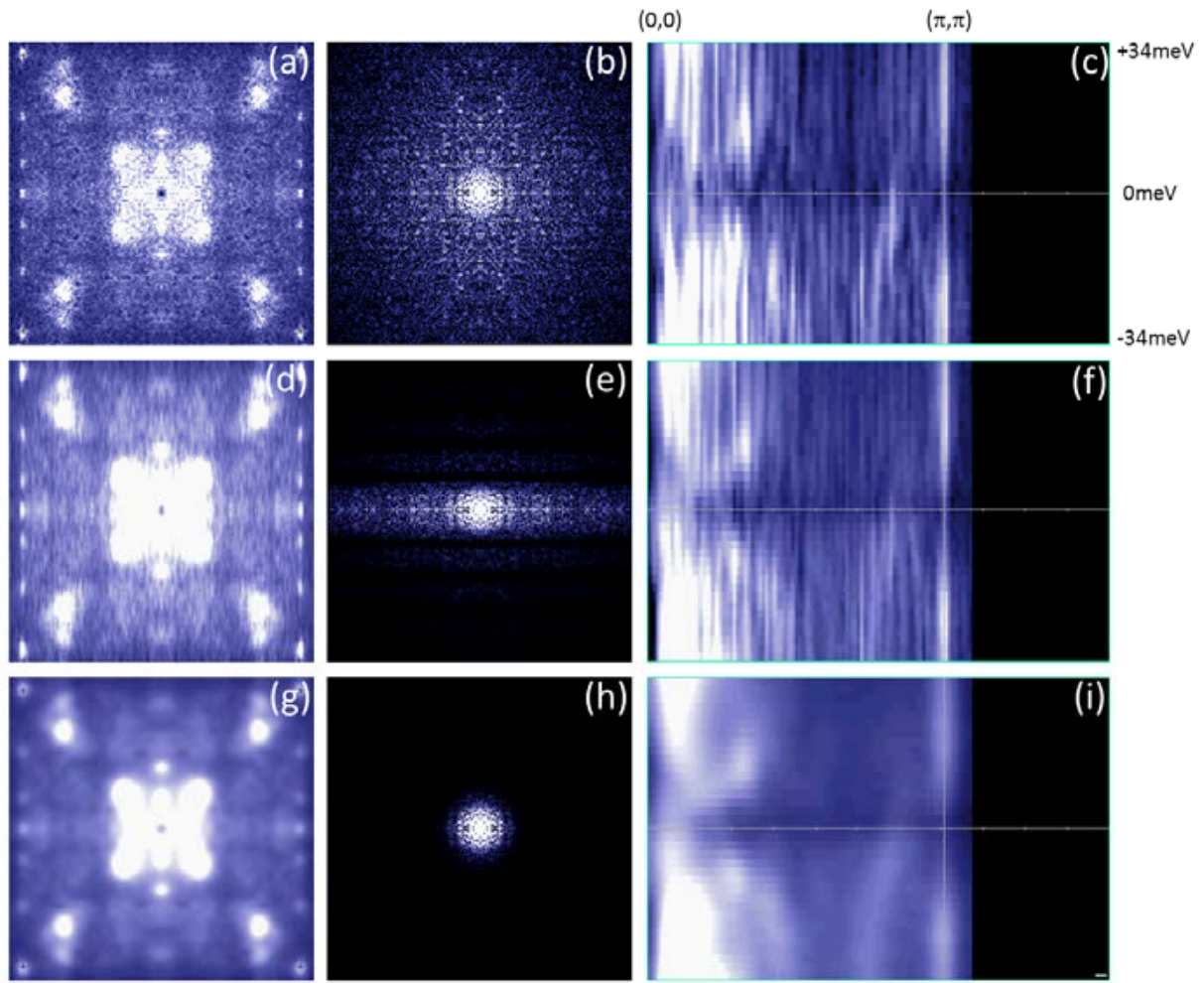

FIG. 1. Comparison of transverse box averaging (conventionally used in peak fitting in high-symmetry axis) and isotropic gaussian averaging with same width. The data is taken on UD37 Bi2212 sample at 4K. (a) Original $q$-space image ($E$=6 meV), (b) its Fourier space amplitude and (c) (0,0)-($\pi$,$\pi$) line-cut. Note the nearly isotropic Fourier space noise distribution in the original $q$-space image. (d) $q$-space image transverse-box-averaged (radius=4 pixel) for (0,0)-($\pi$,$\pi$) line cut, its (e) Fourier space amplitude and (f) (0,0)-($\pi$,$\pi$) line-cut. (g) Isotropic-gaussian-averaged (radius=4) $q$-space image, its (h) Fourier space amplitude and (i) (0,0)-($\pi$,$\pi$) line-cut. Envelopes of (e) and (h) correspond to 1D Sinc and 2D isotropic Gaussian functions, respectively.

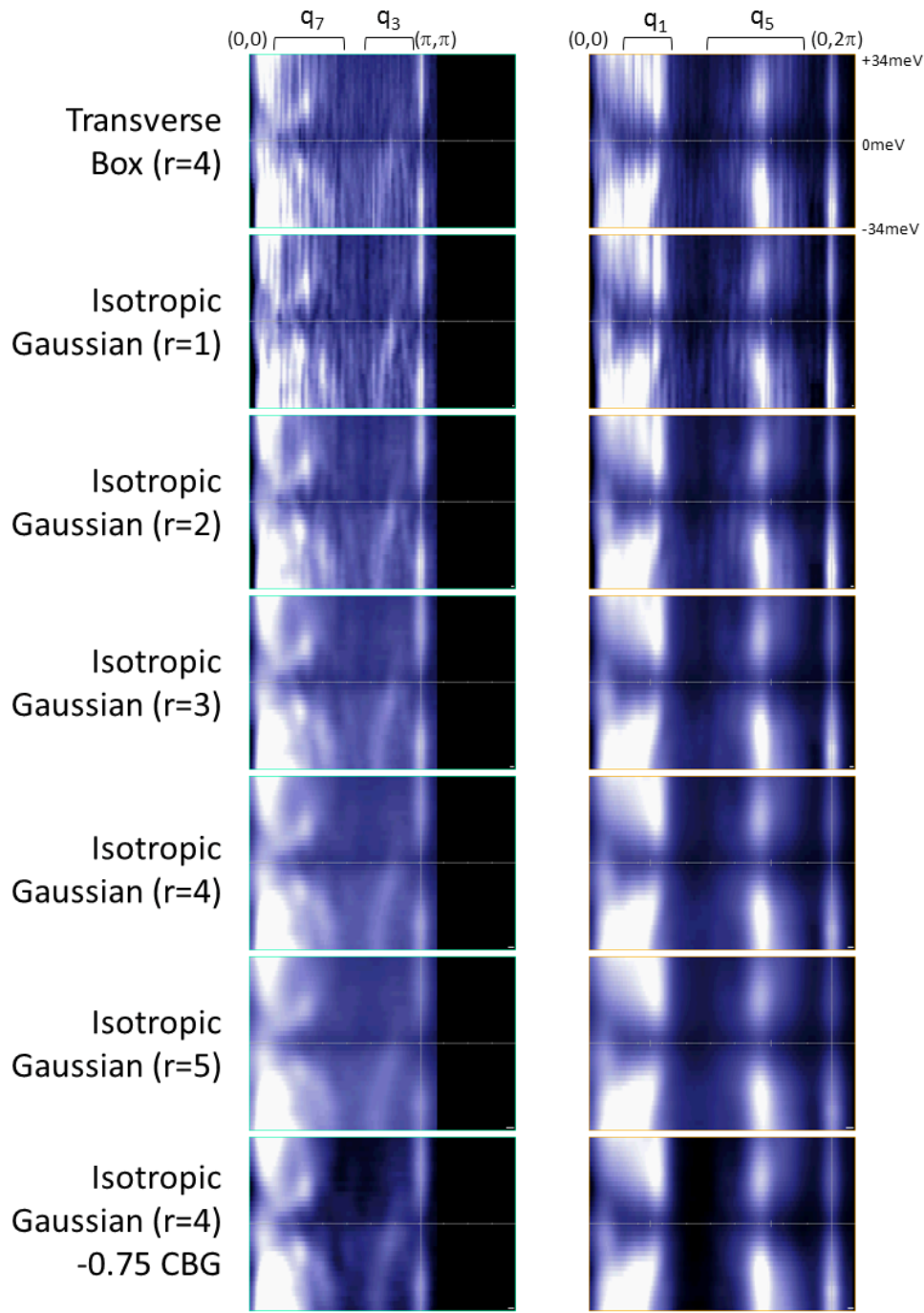

FIG. 2. Effects of IGA radius on FTSTS line cuts. In the last row a background subtraction is applied after IGA4. The data was taken on UD40 Bi2212 at 4K.

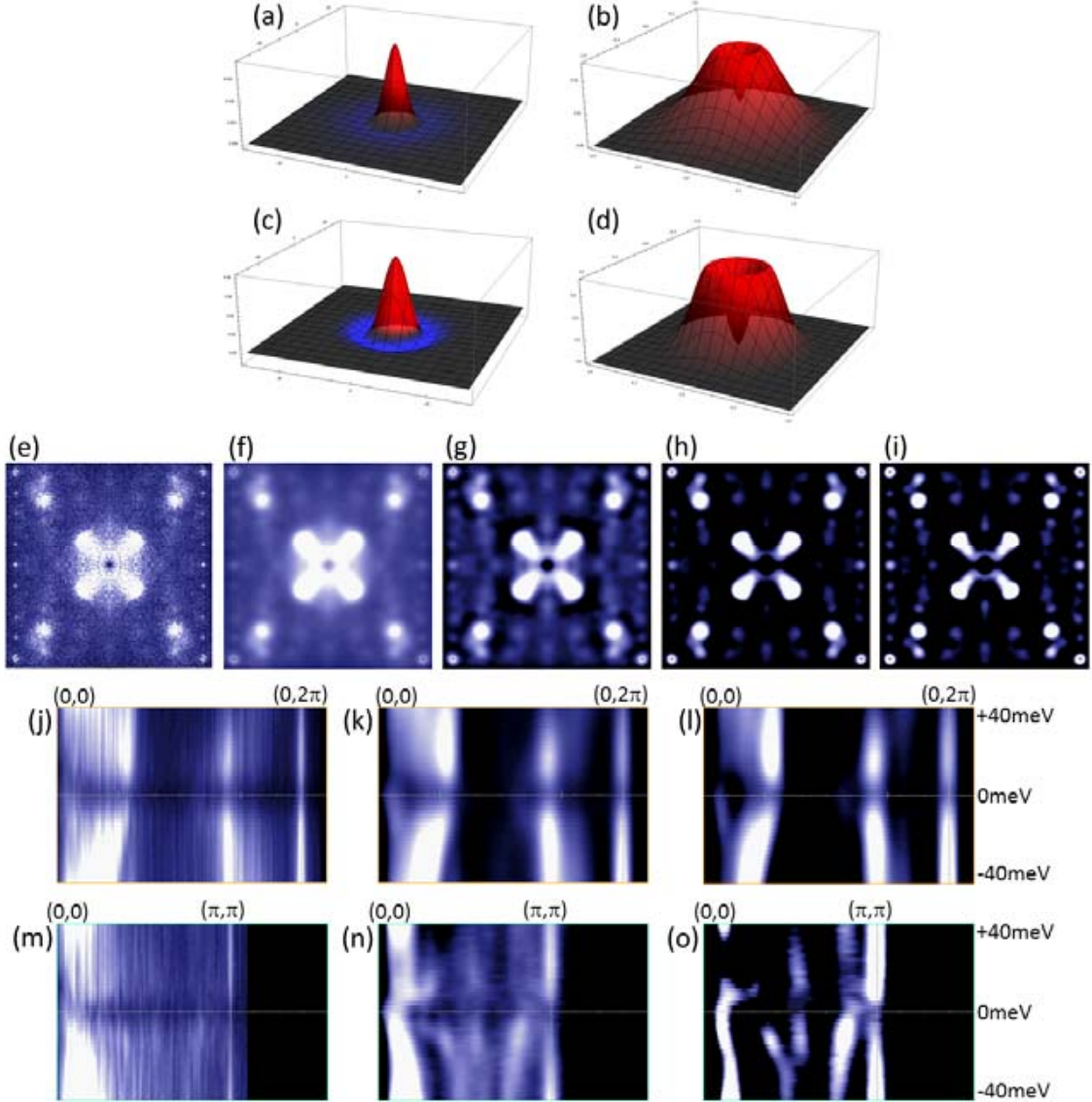

FIG. 3. (Color online) Effects of noise & background removal by isotropic bandpass filters on FTSTS $q$-space images and line-cuts. The data is taken on UD20 Bi2212 at 4K. Convolution kernel of IGA4-0.75X15 (kernel=Gaussian(r=4(pixels))-0.75xGaussian(r=15)) in (a) $q$-space and (b) its Fourier space. Convolution kernel of NL-IGA7 (kernel=-Laplacian(Gaussian($r$=7))) in (c) $q$-space and (d) its Fourier space. In (a) and (c), blue color represents region with negative value. (e) Original FTSTS $g(\mathbf{q},18$ meV) image of UD20 Bi2212 taken at 4K. (f) IGA4. (g) IGA4-0.75X15. (h) IGA4-1.0X15. (i) NL-IGA7. Note the similarity of (h) and (i) due to similar cutoff frequencies and complete DC removal. (j)-(l) Line-cuts of FTSTS $g(\mathbf{q},E)$ of UD20 Bi2212 taken at 4K in (0,0)-(0,2$\pi$) direction. (j) Transverse averaged with 9-pixel width. (k) IGA4-0.75X15. (l) NL-IGA7. (m)-(o) Line-cuts in (0,0)-($\pi,\pi$) direction. (m) Transverse averaged with 9-pixel width. (n) IGA4-0.75X15. (o) NL-IGA7. For (e)-(i), the FTSTS image size is 220x220 pixels and the four crosses at corners indicate ($\pm 2\pi$,0) and (0,$\pm 2\pi$). In every experimental FTSTS plot used in this article, the central peak at (0,0) is suppressed by multiplying (1-gaussian) of 8 pixel radius. In this article IGA4-0.75X15 is used for experimental data and IGA4 is used for simulated data with no noise. Color scale is linear and arbitrary for each panel.

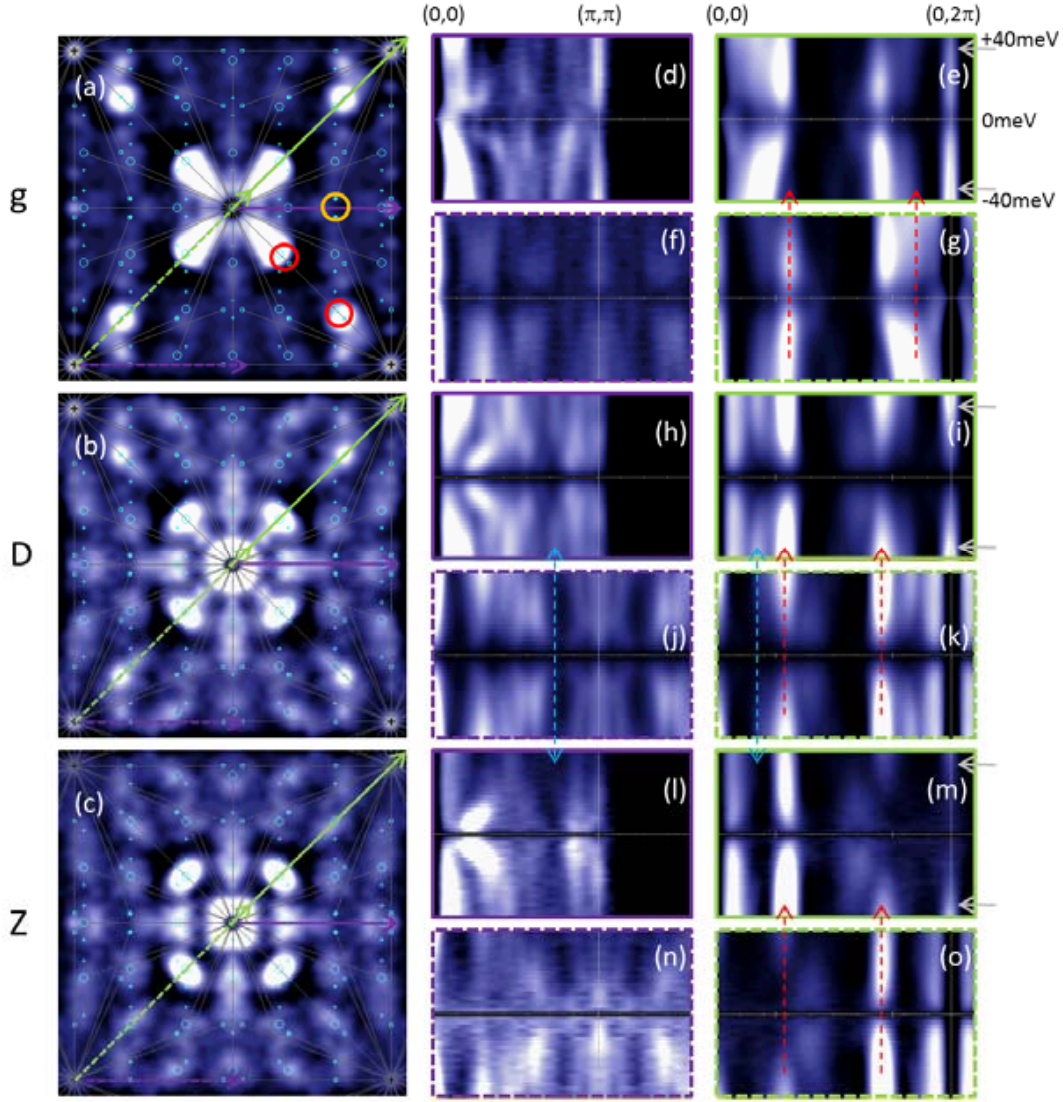

FIG. 4. (Color online) IGA4-0.75X15 (a) $g(\mathbf{q},18$ meV), (b) $D(\mathbf{q},18$ meV) ($D(\mathbf{r},E)\equiv g(\mathbf{r},E)-g(\mathbf{r},-E)$) and (c) $Z(\mathbf{q},18$ meV) with symbols showing the original octet peaks (circles) and the $\mathbf{q}$-vector mixing effect due to the atomic peaks (squares) and the supermodulation peaks (crosses). The data is taken on UD20 Bi2212 at 4K. The strong $\mathbf{q}_1$ and $\mathbf{q}_5$ aligned in the atomic peak direction can in principle cause cross-contamination problem (red circles). Note that $\mathbf{q}_7$ and $\mathbf{q}_3$ have little such problem because $\mathbf{q'}_4$ is weak and mostly off-axis (orange circle). (d)-(o) $g$, $D$ and $Z$ line cuts along (d,h,l) $(0,0)\rightarrow(\pi,\pi)$, (e,i,m) $(0,0)\rightarrow(0,2\pi)$, (f,j,n) $(0,-2\pi)\rightarrow(\pi,-\pi)$ and (g,k,o) $(0,-2\pi)\rightarrow(0,0)$. Comparison of individual features in $D$ and $Z$ can be used in identifying the set-point artifacts in $D$ (blue dashed arrows) and ignoring them in the subsequent analyses. Comparisons between $(0,0)\rightarrow(0,2\pi)$ (green solid arrows and boxes) and $(0,-2\pi)\rightarrow(0,0)$ (green dashed) and between $(0,0)\rightarrow(\pi,\pi)$ (purple solid) and $(0,-2\pi)\rightarrow(\pi,-\pi)$ (purple dashed) can be used in identifying possible $\mathbf{q}$-vector mixing artifacts (red dashed arrows) that can in principle appear even in $Z$ if the atomic peak remains strong at high energy (gray arrows) and the source peak ($\mathbf{q}_1$) is much stronger than the target peak ($\mathbf{q}_5$) as shown in (m).

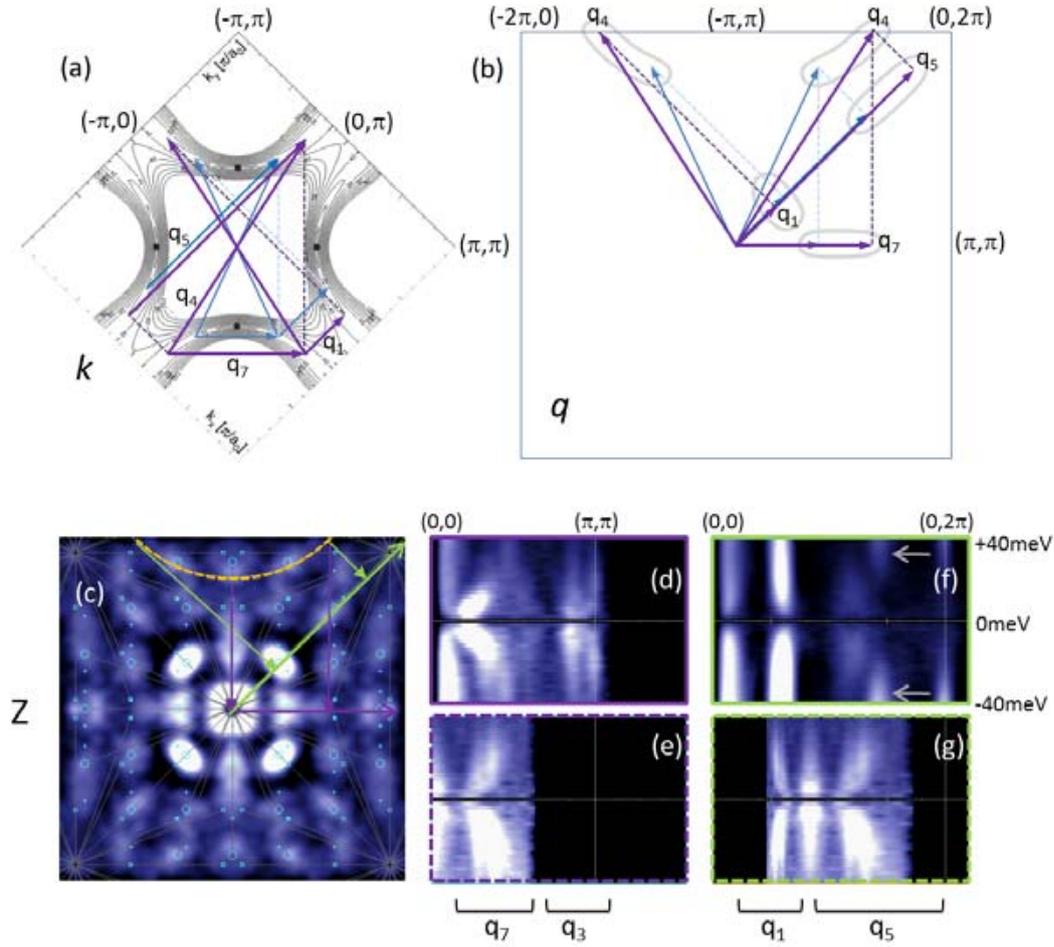

FIG. 5. (Color online) Using projections of $q_4$ curved line cut for octet consistency test. The data is taken on UD20 Bi2212 at 4K. Geometrical relationship between $q_1$, $q_4$, $q_5$, and $q_7$ in (a) $k$-space and (b) $q$-space. Taking a curved line cut along a path (yellow dashed curve) running though the maxima of $q_4$ at all energies and projecting it onto (0,0)-(0,2π) and (0,0)-(π,π) lines can simulate the behavior of $q_1$, $q_5$ and $q_7$. (c) IGA4-0.75X15 of $Z(q,18\text{ meV})$ with yellow dashed curve representing the line cut path for $q_4$ (a circular arc with center at 1.76(-π,π) and radius 1.30π). (d) (0,0)-(π,π) line cut showing $q_7$ and $q_3$ (e) (0,0)-(π,π) projection of $q_4$ line cut simulating $q_7$ (f) (0,0)-(0,2π) line cut showing $q_1$ and $q_5$ (g) (0,0)-(0,2π) line cut simulating $q_1$ and $q_5$. $q_5$ simulated by $q_4$ in (g) shows a single component curve, while the original $q_5$ in (f) shows a heterogeneous component marked by gray arrows that cannot be explained by octet model. It can be either a high energy ECG related peak or an alias of $q_1$ due to the atomic peak. Color scale is linear and arbitrary for each panel.

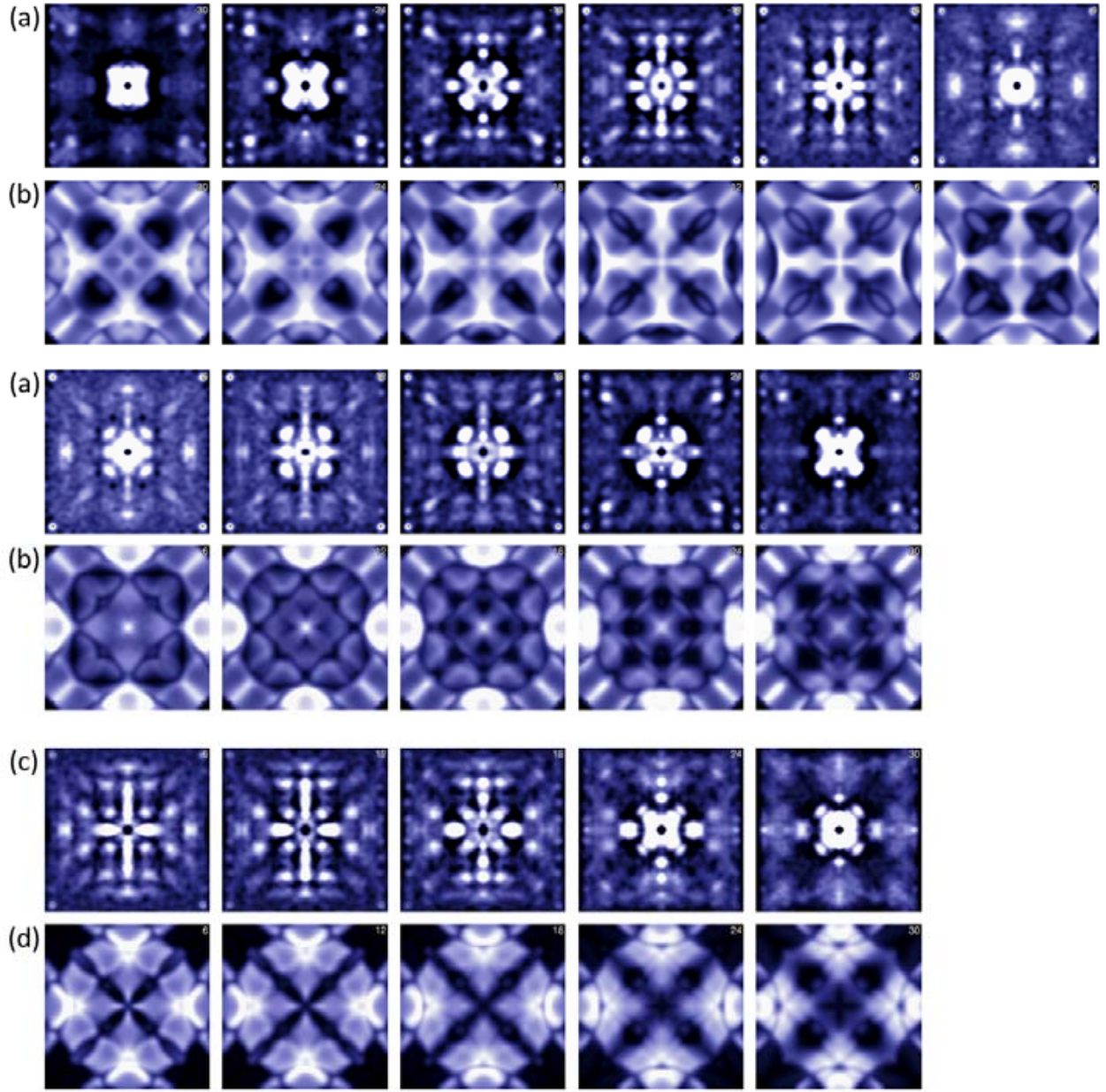

FIG. 6. Comparison between experimental and simulated (non-magnetic scattering) *q*-space constant energy layers using IGA. (a) Experimental IGA4-0.75X15 $g(q,E)$ on UD74 Bi2212 at 4K, (b) Simulated IGA4 $g(\boldsymbol{q},E)$ based on reference [6]. The simulation parameters ($\Delta_0$=0.055 eV, $t_0$=0.0715 eV, $\delta$=0.006 eV, $V_s$=0.1 V and $V_m$=0 V) are the same as those for FIG. 8(e)-(h). (c) Experimental IGA4-0.75X15 $D(\boldsymbol{q},E)$, (d) Simulated IGA4 $D(\boldsymbol{q},E)$. The energy (meV) is indicated in the upper-right corner of each panel. Color scale is linear and arbitrary for each panel.

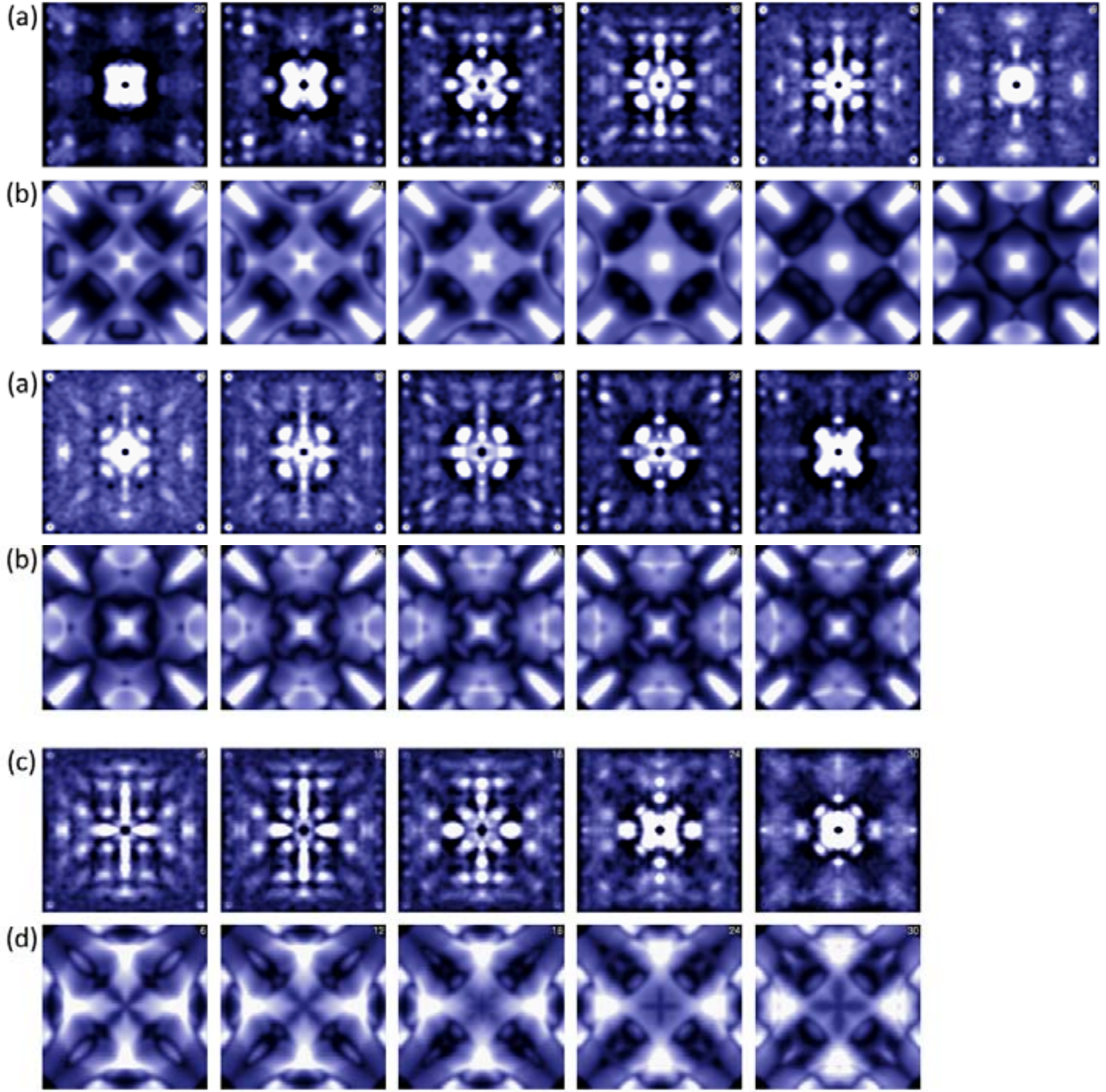

FIG. 7. Comparison between experimental and simulated (magnetic scattering) q-space constant energy layers using HP-FTSTS. (a) Experimental IGA4-0.75X15 $g(\bm{q},E)$ on UD74 Bi2212 at 4K, (b) Simulated IGA4 $g(\bm{q},E)$ based on reference [6]. The simulation parameters ($\Delta_0$=0.055 eV, $t_0$=0.0715 eV, $\delta$=0.006 eV, $V_s$=0 V and $V_m$=0.1 V) are the same as those for FIG. 8(i)-(l). (c) Experimental IGA4-0.75X15 $D(\bm{q},E)$, (d) Simulated IGA4 $D(\bm{q},E)$. The energy (meV) is indicated in the upper-right corner of each panel. Color scale is linear and arbitrary for each panel.

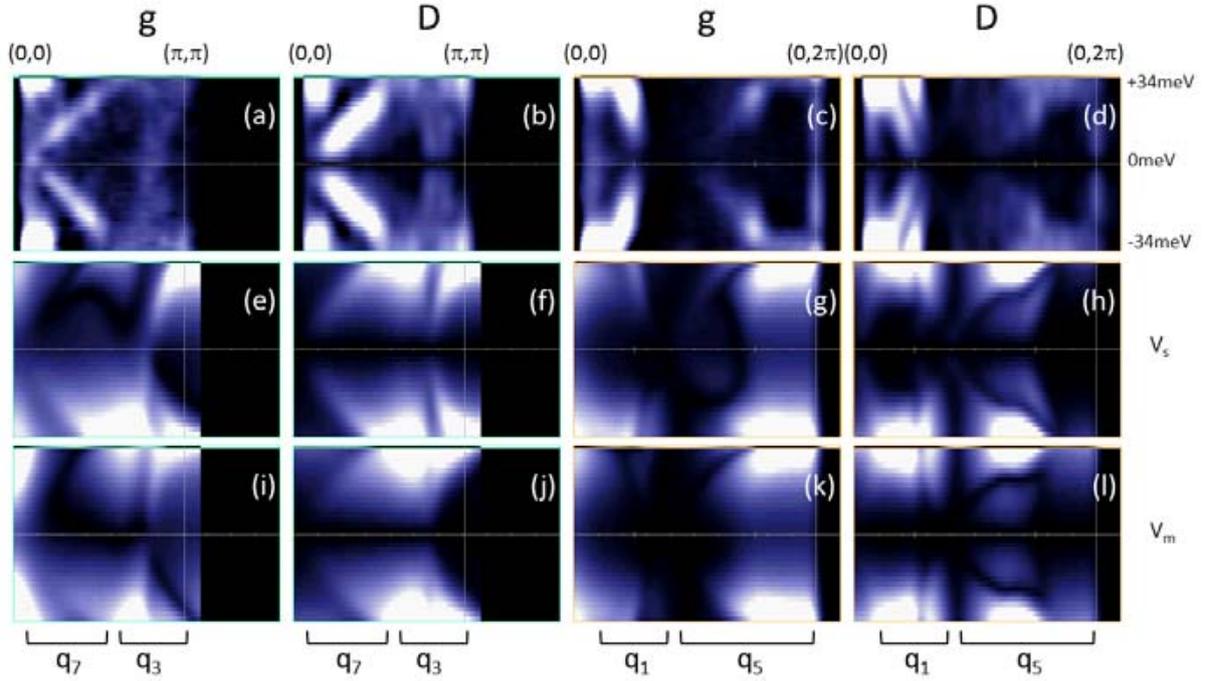

FIG. 8. Comparison between experimental and theoretical FTSTS line cuts with IGA. (a)-(d) Experimental IGA4-0.75X15 $g(\mathbf{q},E)$ and $D(\mathbf{q},E)$ line cuts for UD74 Bi2212 at 4K in $(0,0)$-$(\pi,\pi)$ and $(0,0)$-$(0,2\pi)$ directions. In this lightly underdoped sample, the $\mathbf{q}_1$-$\mathbf{q}_5$ cross-contamination is hardly noticeable even in $g$ and $D$. (e)-(l) Simulated IGA4 $g(\mathbf{q},E)$ and $D(\mathbf{q},E)$ based on reference [6]. The simulation parameters ($\Delta_0$=0.055 eV, $t_0$=0.0715 eV, $\delta$=0.006 eV) are adjusted to reproduce the dispersions for $\mathbf{q}_7$ and $\mathbf{q}_3$ in $g$ and $D$. (e)-(h) Results for non-magnetic impurity scattering ($V_s$=0.1 eV, $V_m$=0 eV). (i)-(l) Results for magnetic impurity scattering ($V_s$=0 eV, $V_m$=0.1 eV). Note the reproduction of the $\mathbf{q}_5$ plateaus at high energies. Color scale is linear and arbitrary for each panel.

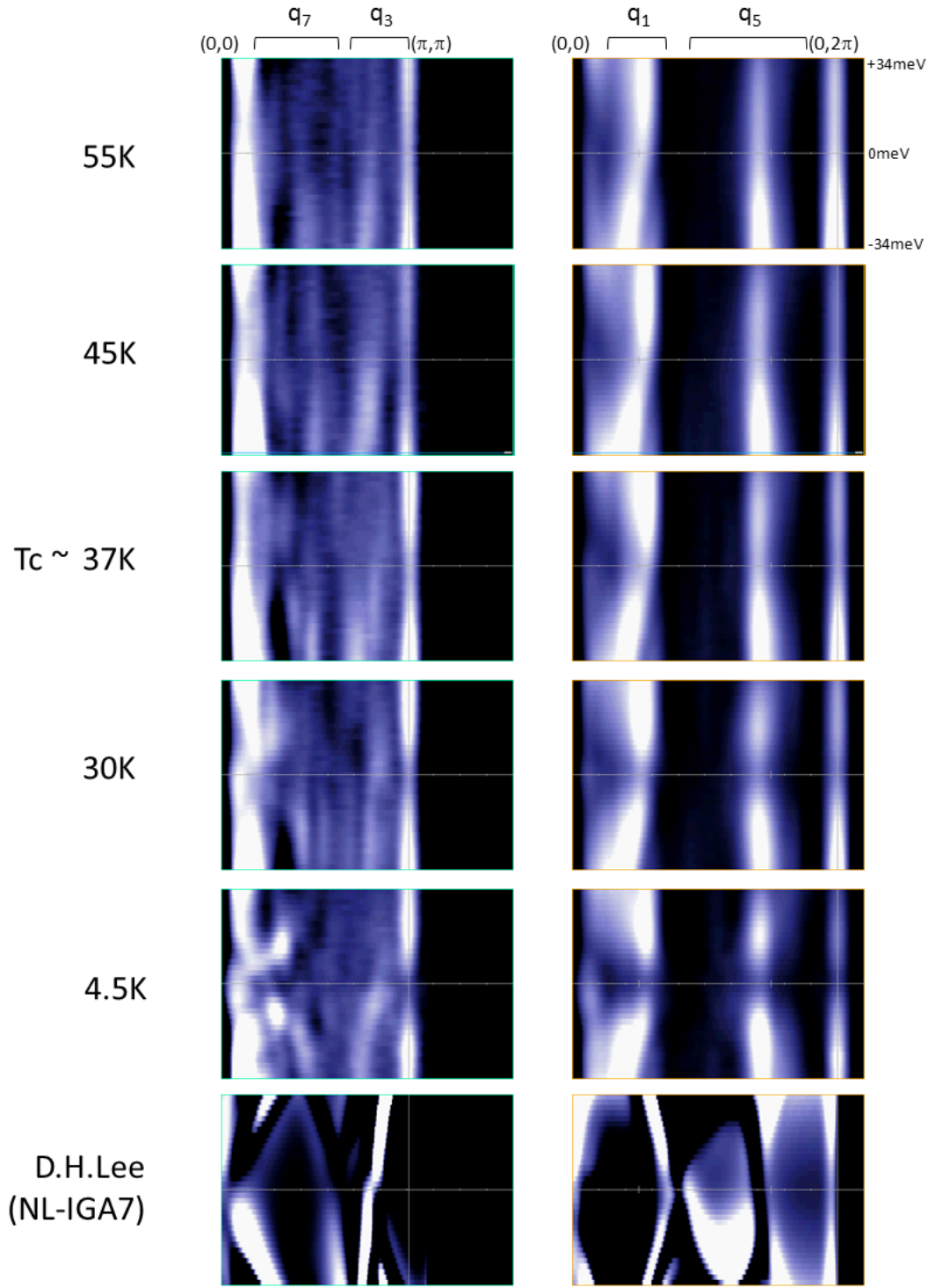

FIG. 9. Temperature-dependent quasiparticle interference of UD37 Bi2212. The experimental data (top 5 rows) shown are cross-sections of $g(r,E)$ of experimental data from reference [4]. Note the increasing asymmetry in $q_1$ dispersion as temperature increases. Fine structures in 4.5K $(0,0)$-$(\pi,\pi)$ line cut are roughly reproduced by a simulation (the bottom row) following the reference [6]. ($\Delta_0$=0.1 eV, $t_0$=0.0615 eV (Norman '94), $\delta$=0.006 eV, $V_s$=0.1 V, $V_m$=0 V). In the simulation, due to the negative-laplacian (NL) processes following the IGA7, the peak dispersions are strongly enhanced.

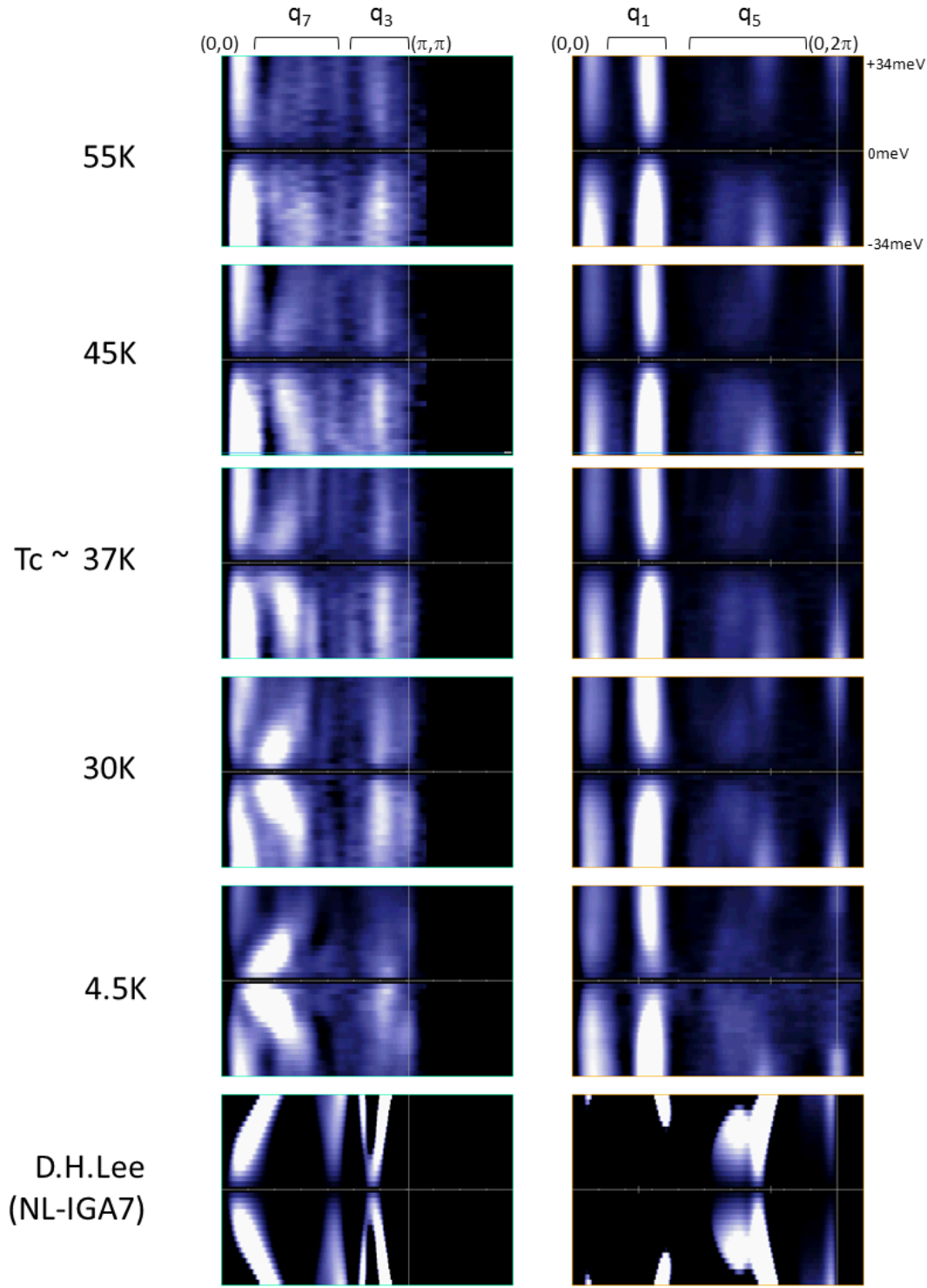

FIG. 10. Temperature-dependent quasiparticle interference of UD37 Bi2212. The experimental data (top 5 rows) shown are cross-sections of $g(\mathbf{r},E)$ of experimental data from reference [4]. Fine structures in 4K $(0,0)$-$(\pi,\pi)$ line cut are roughly reproduced by a simulation (the bottom row) following the reference [6]. ($\Delta_0$=0.1 eV, $t_0$=0.0615 eV (Norman '94), $\delta$=0.006 eV, $V_s$=0.1 V, $V_m$=0 V). In the simulation, due to the negative-laplacian (NL) processes following the IGA7, the peak dispersions are strongly enhanced.

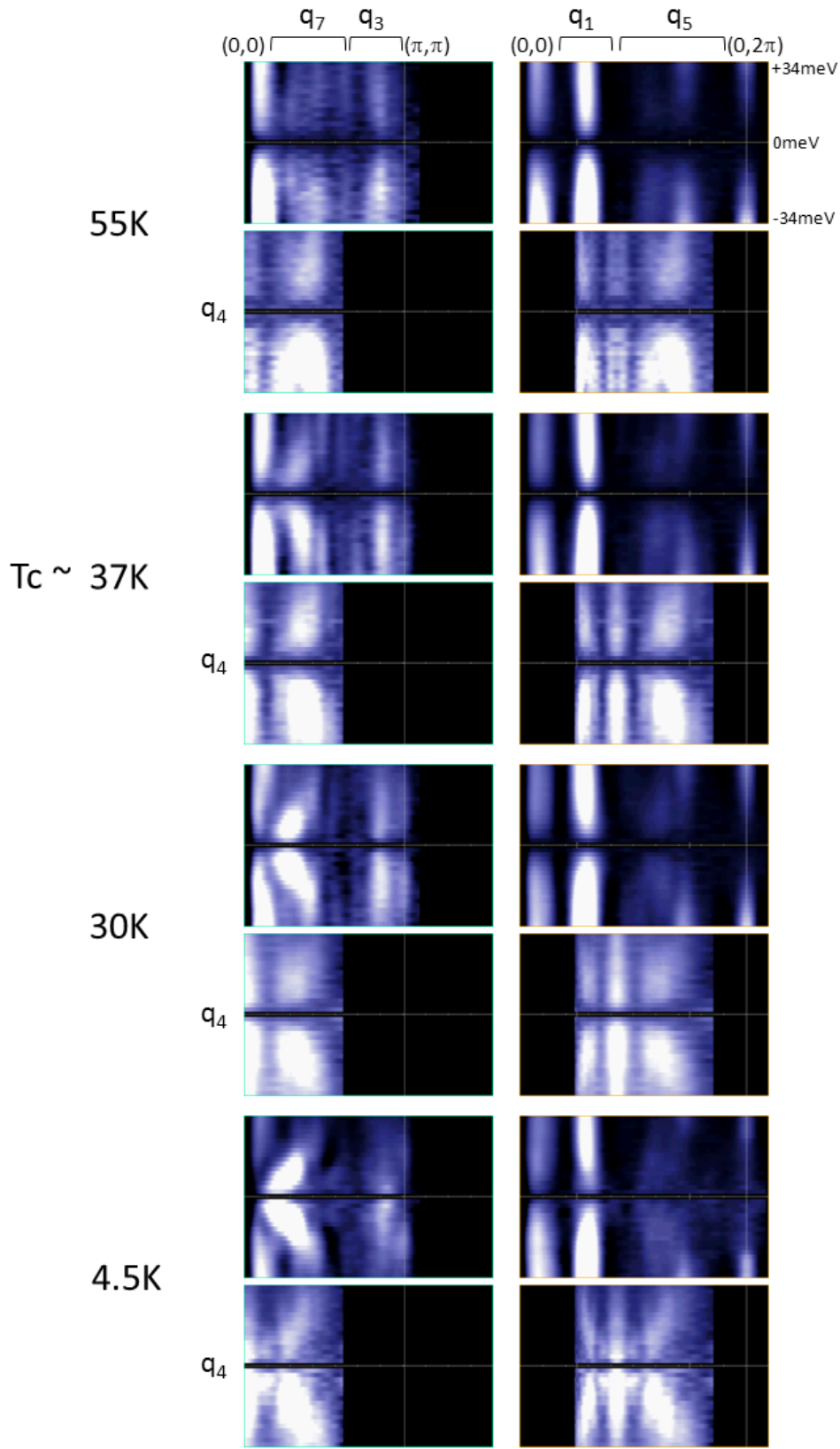

FIG. S1. Original *Z* line cuts and projected (per FIG.5) $q_4$ *Z* line cuts. The data was taken on UD37 Bi2212 at 4K.

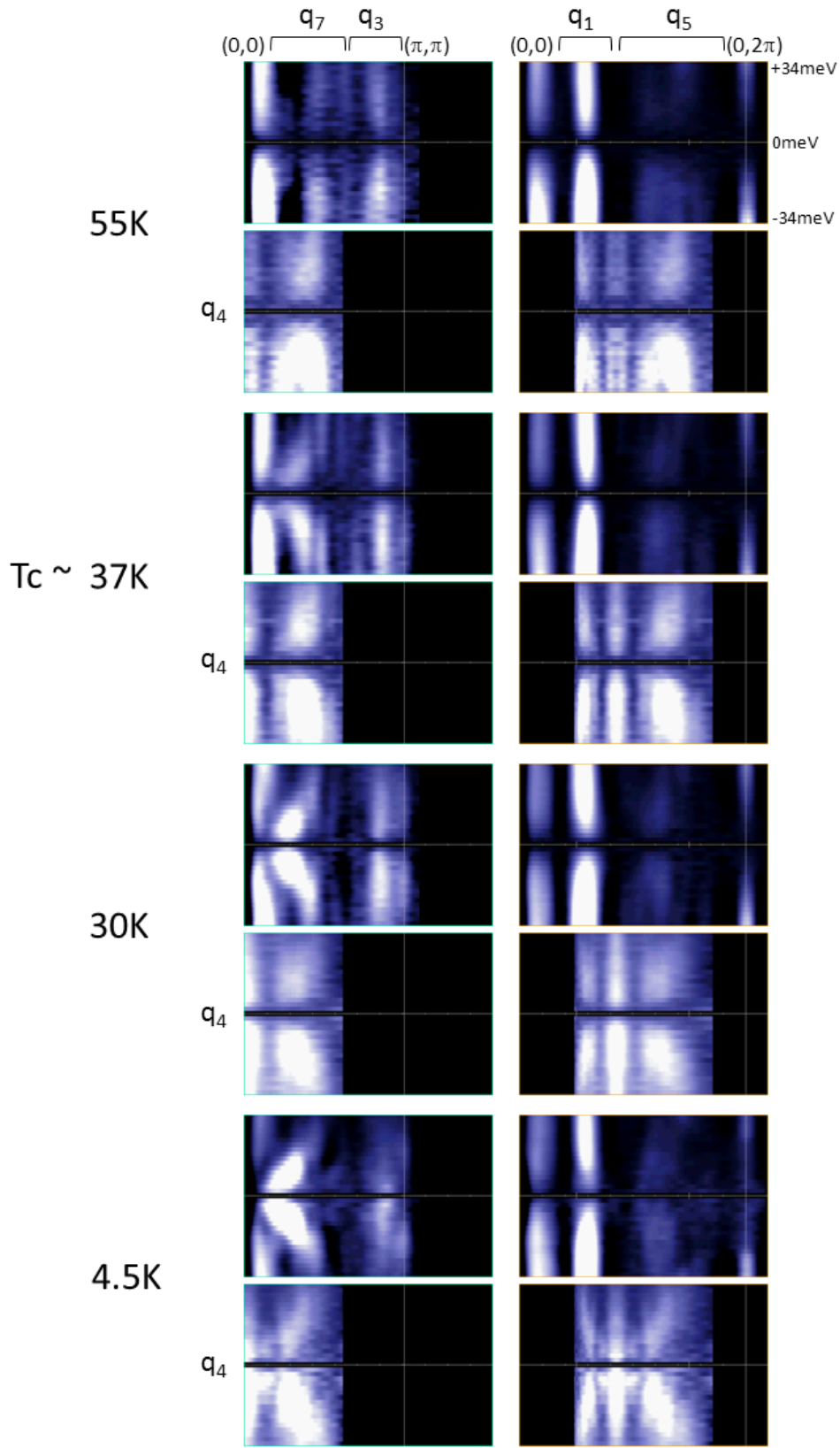

FIG. S2. Alias(by Bragg peaks and supermodulation peaks)-compensated $Z$ line cuts and projected $q_4$ $Z$ line cuts. By the comparison with FIG S1, the removal of the artifact $q_5$ component with $q_5 \sim 2\pi - q_1$ is demonstrated. The data was taken on UD37 Bi2212 at 4K.